\documentclass[aip,preprint,letterpaper,groupedaddress,longbibliography]{revtex4-2}
\usepackage{amsmath,amsthm,amsfonts}
\theoremstyle{definition}

\usepackage{multirow}
\usepackage{array}
\usepackage{soul}
\usepackage{float}
\usepackage{graphicx}
\usepackage{afterpage}
\usepackage{tabularx}
\usepackage{siunitx}

\usepackage[font=small,labelsep=quad,format=plain]{caption}
\usepackage{subcaption}
\usepackage[ colorlinks=true,citecolor=blue,linkcolor=blue,urlcolor=blue]{hyperref}
\usepackage{cleveref}
\crefname{equation}{Eq.}{Eqs.}
\crefname{section}{Sec.}{Secs.}
\crefname{subsection}{Sec.}{Secs.}
\crefname{appendix}{Appendix}{Appendices}
\crefname{figure}{Fig.}{Figs.}
\crefname{table}{Table}{Tables}
\crefname{proposition}{}{}
\crefname{corollary}{}{}

\usepackage{bm}
\usepackage[dvipsnames]{xcolor} 

\begin{document}

\title{Characterization of ultrathin nickel films deposited by thermal laser evaporation}

\author{David S. Catherall}
\affiliation{Division of Engineering and Applied Science, California Institute of Technology, Pasadena, CA 91125, USA}
\author{Yifei Yan}
\affiliation{Division of Engineering and Applied Science, California Institute of Technology, Pasadena, CA 91125, USA}
\author{Finley B. Donachie}
\affiliation{Division of Engineering and Applied Science, California Institute of Technology, Pasadena, CA 91125, USA}
\author{Azmain A. Hossain}
\affiliation{Division of Engineering and Applied Science, California Institute of Technology, Pasadena, CA 91125, USA}
\author{Austin J. Minnich}
\email{aminnich@caltech.edu}
\affiliation{Division of Engineering and Applied Science, California Institute of Technology, Pasadena, CA 91125, USA}
\date{\today}

\begin{abstract}
Thermal laser evaporation is a physical vapor deposition technique of increasing interest because of its ability to evaporate essentially any solid element, even the most refractory such as W. However, many films deposited by this method, especially non-epitaxial films, remain to be characterized; further, key system components such as the laser delivery system have not been described in detail. Here, we present the evaporation and characterization of ultrathin Ni films deposited with a home-built thermal laser evaporation system. The system employs a continuous-wave 1 kW fiber laser (1070 nm) focused to sub-millimeter diameter onto a Ni target rod mounted inside an ultrahigh-vacuum chamber. The laser heats the target to a temperature high enough to produce vapor for film deposition; for Ni, this temperature is around the melting point of 1725 K. We report the characterization of the surface roughness, composition, and room-temperature electrical properties of the films along with the design of the major components of our system. This work advances the growing consensus regarding the potential of thermal laser evaporation for thin film deposition and epitaxy and provides the necessary design information to facilitate broader adoption of the technique.
\end{abstract}

\maketitle
\newpage

Physical vapor deposition (PVD) of thin films is performed by various well-established techniques, including thermal and electron beam evaporation (e-beam), sputtering, pulsed laser deposition (PLD), and others. \cite{mattox2010pvd} Evaporative methods rely on the condensation of vapor created by heating of source material onto a substrate. Sputtering and PLD are non-equilibrium techniques based on the ejection of material from a target by plasma ion impingement and the ablation of a target with energetic laser pulses, respectively. These techniques enable directional or conformal deposition of diverse elemental and compound materials.

Film deposition by evaporation is in routine use, for instance for deposition of metals for devices and for epitaxial growth using molecular beam epitaxy (MBE). However, evaporation of refractory elements is a long-standing challenge owing to their low vapor pressure. For instance, elements like W, Ru, Nb, Ta, and Mo cannot be deposited using effusion cells, and  while e-beam evaporation of these elements is technically possible, the deposition rates are usually well below 1 \AA/sec (typical e-beam evaporation rates are between 1 and 50 \AA/sec). Additionally, careful monitoring and preparation are required to avoid damaging the evaporation system due to the required high powers and temperatures.

The use of continuous-wave lasers to thermally evaporate source material for deposition has potential to overcome these limitations.  \cite{Groh:1968, Cheung:1988, Sankur:1988, Braun:2019} In recent years, this technique has been termed thermal laser evaporation  \cite{Braun:2019,Smart:2021,Kim:2021_1,Kim:2022} (TLE) following earlier works which termed it variations of laser evaporation \cite{Ban:1970,Fujimori:1980,Sankur:1983,Sankur:1985,Brauns:1986,Cheung:1988,YIN:1997,MINETA:1990,Jasti:2024}. The technique has also been called thermal laser epitaxy \cite{AlTaleb:2022,Smart:2023,Smart:2023_2,Kim:2023,Smart:2024,Majer:2024_1,Majer:2024_2,Majer:2024_3,Smart:2025}, which is also abbreviated TLE. In this work, we take TLE to mean thermal laser evaporation as the deposited films here are non-epitaxial. TLE uses continuous-wave (CW) lasers focused onto freestanding targets (slugs or crucibles) to provide the necessary thermal energy to produce vapor for deposition. It differs from PLD (or laser-MBE) in that heating is performed using CW rather than pulsed lasers, leading to deposition by thermal evaporation rather than an ablation process. \cite{Gross-laserMBE, Lowndes-PLD-1996, Cheung:1988} TLE also may incorporate laser-based substrate heating, which can achieve temperatures up to the melting point of the substrate and as such is well suited for high temperature growth and \textit{in-situ} annealing. \cite{Hanna:2025,Kim:2025,Majer:2024_3,Majer:2024_2,Majer:2024_1,Smart:2024,Kim:2023,AlTaleb:2022}.

The quasi-stationary nature of deposition by TLE makes the deposition process more similar to thermal PVD techniques like e-beam evaporation but with several advantages. TLE requires no crucible for many materials, allowing for the deposition of both refractory metals and elements that may react with a crucible. The lack of a crucible also permits higher source temperatures which would normally damage a crucible or induce contamination, meaning that refractory elements can be deposited at rates more typical in PVD systems (over 1 \AA/sec). Because the energy source is located outside of vacuum, the technique is especially suitable for reactive evaporation since there are no hot filaments or resistive heaters that could be degraded by exposure to process gases (e.g. O\textsubscript{2}) \cite{Sankur:1988}. Additionally, the source-substrate distance is typically an order of magnitude smaller than in e-beam evaporation or MBE, allowing for higher pressures of reactive background gases without scattering the metallic vapors. 


TLE was initially developed contemporaneously with MBE, electron-beam evaporation, and PLD in the late 1960s. A variety of oxides and compounds were deposited using TLE \cite{Cheung:1988, Groh:1968}, as well as pure elements including C \cite{Fujimori:1980,FUJIMORI1:1982}, Os \cite{Maier:1979}, Pt \cite{Hess:1972}, and Te \cite{Hass:1969}. However, laser technology of the time was inadequate for TLE to compete with other techniques due to two difficulties. First, the most practical CW laser of the time was the CO\textsubscript{2} laser, which was then limited in power to a few hundred watts \cite{Hass:1969}. Further, its 10.6 \unit{\um} wavelength is not well absorbed by metals, leading many investigators to focus on oxide or compound targets \cite{Cheung:1988,Duley:1983}. Second, the coating of optical elements exposed to the heated target added complexity to the energy delivery system compared to effusion cells or electron beam systems \cite{Sankur:1988}. Although this latter drawback could be partially worked around by using mirrors and apertures \cite{Sankur:1988,Sankur:1985,Hass:1969}, the other techniques were technically simpler to implement.

Due in large part to the maturation of other PVD techniques, TLE was neglected for decades. Recent developments in laser technology, however, have made the technique increasingly viable. The key development has been the commercial availability of high-power fiber lasers, which emit at micron wavelengths and feature output powers up to and exceeding several kilowatts. Fiber lasers are more useful than the CO\textsubscript{2} laser for evaporating most elements due to the order of magnitude higher absorption of metals at around 1 micron compared to 10.6 microns \cite{Duley:1983}. Recently, a group at the Max Planck Institute at Stuttgart has developed a modern implementation of TLE using high-power CW fiber lasers and deposited many elements, oxides, and nitrides. \cite{Braun:2019, Smart:2021, Kim:2021_1, Kim:2022,Smart:2023_2, Kim:2025} However, modern TLE has only been performed by one group to date, and many of the details of key system components such as the beam delivery system have not been reported. Further, although the deposition of nearly all solid, nonradioactive elements has been reported, the characterization of the non-epitaxial films has been limited to only a few cross-sectional SEM images \cite{Braun:2019,Smart:2021,Smart:2023_2}. Whether TLE can produce non-epitaxial films which meet or exceed the characteristics of those deposited by traditional PVD techniques such as sputtering and electron beam evaporation has not yet been determined.

Here, we report the deposition of ultrathin Ni films using a home-built TLE system employing a 1 kW CW fiber laser with a 1070 nm wavelength. A deposition rate of 0.35 \AA/sec was achieved with 177 W of CW laser power. The deposited film was $14.7\pm0.1$ nm thick and exhibited RMS roughness $R_q = 1.10\pm0.14$ nm. The electrical resistivity, known to be a sensitive probe of the quality of electrically-conducting films, was measured to be $22\pm0.2$ $\mu \Omega$ cm, in quantitative agreement with the best prior value reported for Ni films of a similar thickness. We also provide an overview of the system design and discuss design considerations for key system components such as the beam delivery system. This work indicates that TLE-deposited films are of equal quality to the best films made by other evaporation methods, and the provided system design details will enable broader adoption of the technique, advancing the science and practice of thin film deposition.

\begin{figure}
    \centering{
        \phantomsubcaption\label{fig:system}
        \phantomsubcaption\label{fig:cutaway}
        \phantomsubcaption\label{fig:optics}
        \includegraphics[width=\textwidth]{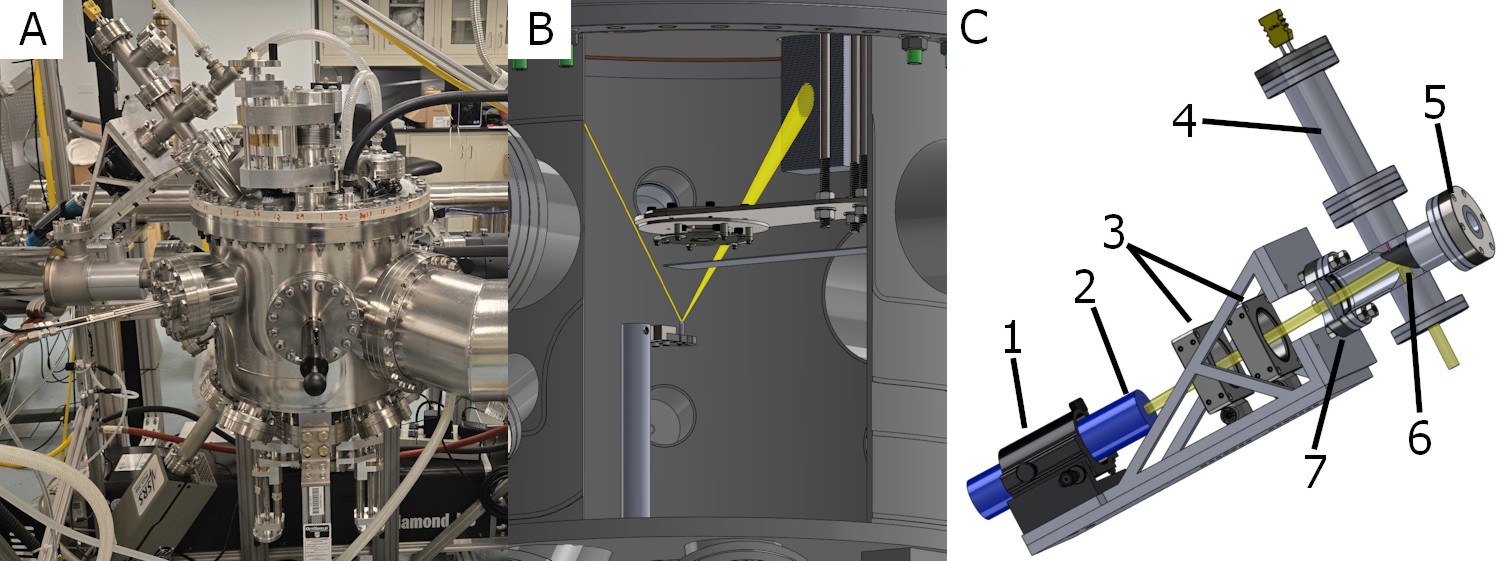}}
    \caption{(A) Photograph of the TLE system at Caltech. At the top left is the laser delivery system, and to its right is the bellows mounted to the substrate holder. Around the center from the left to right are a RHEED system, QCM, shutter control, and turbopump (not shown) respectively. At bottom are the RGA and two bellows to which the target holders are mounted. (B) Cut-away rendering of the system interior from the position of the QCM. The laser is shown in yellow, specularly reflecting off the target onto the beam dump. Above the target are the shutter and substrate, suspended from its holder. (C) Rendering of the fiber laser delivery system. 1. Kinematic laser mount 2. Laser collimator 3. Lenses 4. Thermocouple 5. Water cooled mirror mount 6. Mirror 7. Laser window.}
    \label{fig:system_optics}
\end{figure}

We begin by presenting the details of our home-built TLE system. A photograph of the system is shown in \cref{fig:system}. It consists of a water-cooled vacuum chamber (Kurt J. Lesker Hydra~Cool™) with a turbopump providing a base pressure below $10^{-9}$ mbar. Inside the chamber are a target holder and substrate holder, which are both mounted on bellows to enable target alignment and sample transfer. The relative positioning is shown in \cref{fig:cutaway}. The chamber is also equipped with a deposition shutter, quartz-crystal microbalance (Telemark) for deposition rate monitoring and a residual gas analyzer (SRS RGA200) to monitor the vacuum conditions.

The laser delivery system is shown in \cref{fig:optics}. The laser is a 1 kW fiber laser (IPG Photonics, YLR-1000-WC) and terminates in a water-cooled free-space coupler and collimator (IPG, D30 F100 AC HLC-8), which is mounted in a kinematic laser mount to enable fine beam alignment. Two C-coated lenses (Thorlabs LA1301-C-ML and LC1093-C-ML) are used to focus the beam to sub-millimeter diameter on the target, which is slightly smaller than that used in other TLE works (c.f. 1.1 mm \cite{Smart:2023_2}). This beam diameter was selected to ensure that the melt pool is contained within the target, ideally avoiding unpredictable angular variations of the specularly reflected beam. It is known that a large melt pool results in a convex surface due to surface tension, \cite{Smart:2023_2} which could specularly reflect the beam in an unpredictable direction and potentially damage system components. While a smaller melt pool might also result in a non-planar surface and a similarly unpredictable reflection, this option was ultimately selected. An antireflective window is used to pass the beam into vacuum (Lesker VPZL-275LYAG). The final optical element is an aluminum off-axis parabolic mirror (Edmund Optics 35-590) mounted on a water-cooled rotatable flange. Pt paste (Tanaka Kikinzoku TR-7091T) is used to enhance thermal contact between the flange and the mirror. A type K thermocouple is placed in contact with the mirror to monitor its temperature. A water cooled beam dump is placed in the path of the beam specularly reflected by the target. 

With this optical configuration, it was found that the deposition could be performed for around 45 minutes while limiting the mirror temperature to below 100~\unit{\celsius}. The mirror temperature steadily increased with time due to thermal resistance from the mirror surface to the water-cooled mounting flange and an increase in film absorbance due to deposition from the target. These issues can be mitigated by water-cooling the backside of a thin ($\sim$2 - 4 mm) mirror and use of a pinhole aperture to minimize deposition onto the mirror. These modifications are the focus of ongoing work but were not used in this study.

\begin{figure}
    \centering{
        \phantomsubcaption\label{fig:target_rod}
        \phantomsubcaption\label{fig:target_cantilever}
        \phantomsubcaption\label{fig:substrate}
        \includegraphics[width=5in]{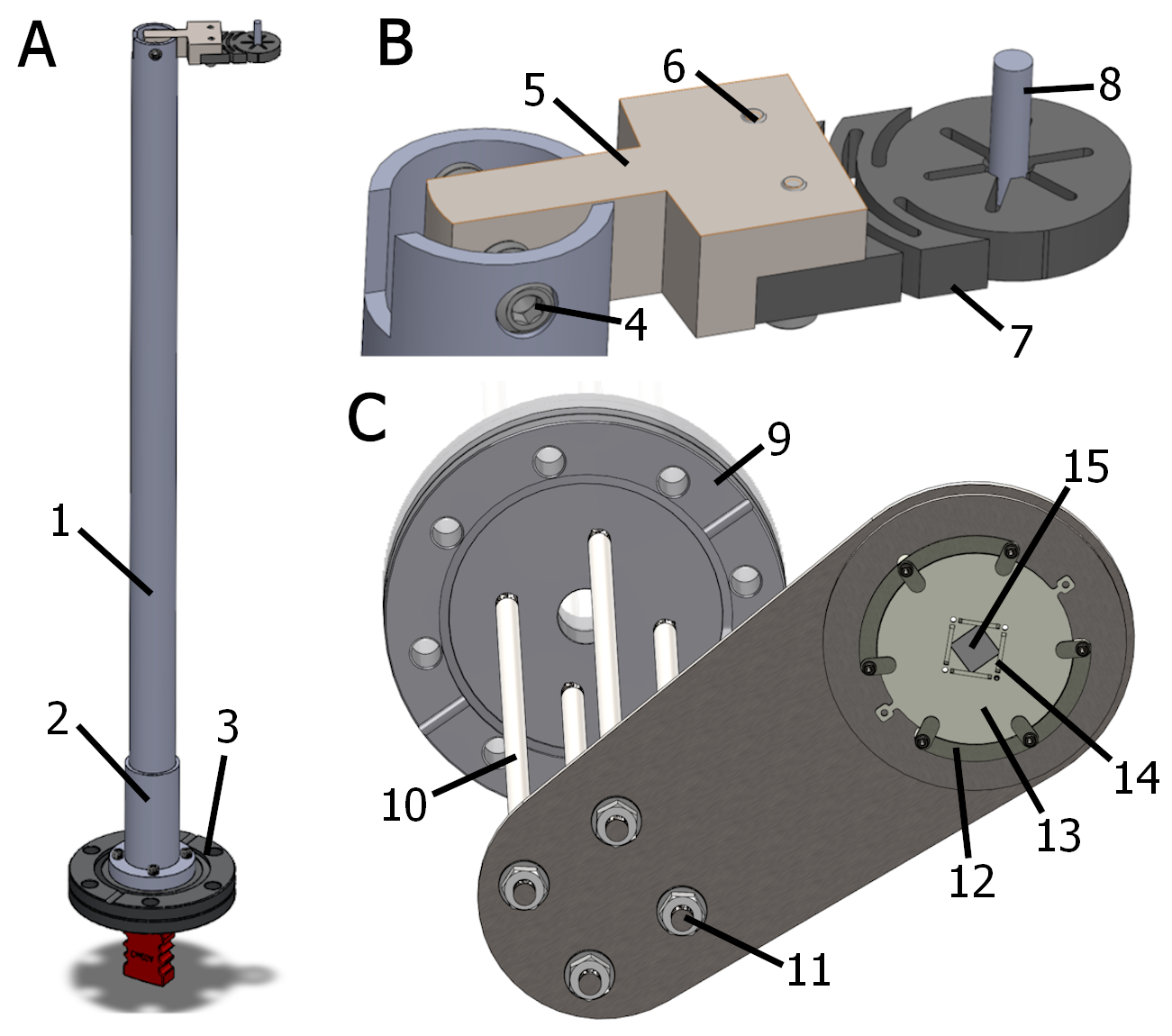}}
    \caption{(A) Target holder apparatus, with thermocouple routed within the interior. 1. Steel tube 2. Aluminum flange adapter 3. C-type thermocouple feedthrough. (B) Target holder cantilever, with thermocouple routed underneath. 4. Titanium screw used as a hinge 5. Steel cantilever support 6. Titanium screws 7. Molybdenum cantilever, with cut outs to increase conduction path between target and screws, and a press-fit hole drilled for the target 8. Target. (C) Substrate holder. 9. Conflat flange 10. Steel rods 11. Nuts and washers to mount the cantilever 12. High-temperature steel fingers 13. Molybdenum substrate disk 14. Molybdenum strips 15. Substrate chip.}
    \label{fig:Holders}
\end{figure}

The target and its mounting structure are shown in \cref{fig:target_rod,fig:target_cantilever}. The target in this work consists of a 12.7 mm long, 6.35 mm diameter Ni rod (Lesker, 99.995\%) which was machined down to 3.13 mm at the base to minimize the thermal conductance to the target holder. We chose Ni for this work to facilitate comparison of electrical data in the literature and because it is less technically challenging to evaporate compared to the refractory elements. Although Ni was primarily chosen for experimental convenience, our findings may be applicable to other metal films, including refractory elements, for which thermal evaporation remains difficult. TLE of refractory elements using our system will be reported in a future publication. 

The target is mounted in a press-fit slot within a molybdenum cantilever, and a type-C thermocouple is mounted under the target to monitor the temperature. This cantilever design was chosen to allow for target insertion through a DN40CF flange. We note that without active cooling, the cantilever can become hot enough during deposition that oxidation would be expected if reactive gases such as O\textsubscript{2} were present. In the present case, deposition was performed in vacuum so that this issue is irrelevant; however, future designs will incorporate water cooling of the target holder.

The substrate and its holder are shown in \cref{fig:substrate}. A $1 \times 1$ cm\textsuperscript{2} sapphire substrate (University Wafer, C-plane, SSP) is mounted on a home-built molybdenum disk that suspends the sample upside down $\sim 70$ mm above the target. Prior literature used a working distance of $\sim 60$ mm,\cite{Braun:2019,Kim:2021_1,Smart:2021,Smart:2025} and as such we expect approximately a 25\% decreased flux, assuming all other parameters are the same. The disk is suspended by steel fingers that are mounted to a cantilever suspended by steel rods from a flange fitted onto an adjustable bellows. Although this design is attractive for simplicity, it does not permit sample rotation. Future implementations will employ a fully concentric design to facilitate sample rotation. The substrate was cleaned by rinsing with acetone, isopropanol, and deionized water. Substrate heating was not performed in this study, but it is under active development using CO\textsubscript{2} laser heating as discussed in Ref.~\onlinecite{Braun:2020}. It is expected that substrate heating will result in higher crystallinity and lower O and C contamination.

The deposition procedure consisted of the following steps. First, the system was pumped down to a base pressure below $10^{-9}$ mbar. No chamber bakeout was performed. A Ti shutter on a linear translation stage covered the sample for the entire process until deposition was initiated. The laser power was held between 119 and 144 W for approximately an hour to bake out adsorbed gases from the target and its holder. After the pressure stabilized around $10^{-7}$ mbar, the power was steadily increased until a steady-state deposition rate was measured in the QCM. The shutter was then opened to allow deposition. For the present study, the deposition rate was 0.35 \AA/sec.

\begin{figure}
    \centering{
        \phantomsubcaption\label{fig:ni-before}
        \phantomsubcaption\label{fig:ni-after}
        \phantomsubcaption\label{fig:Deprate}
        \phantomsubcaption\label{fig:Fluctuations}
        \includegraphics[width=\textwidth]{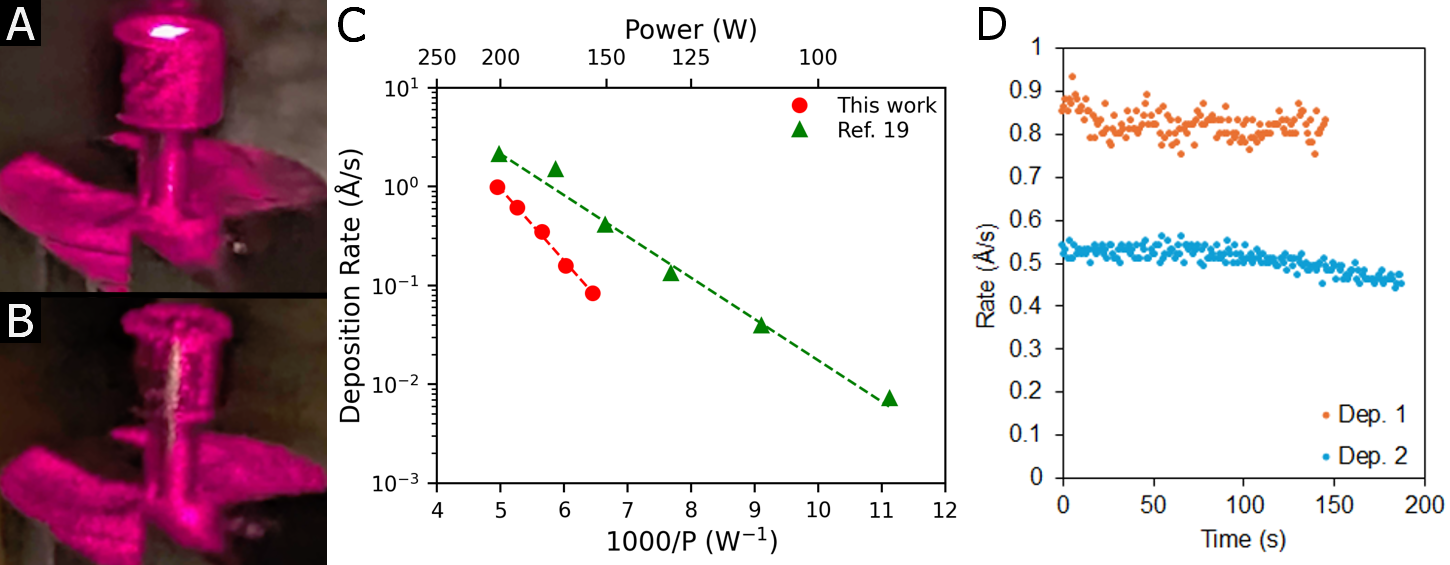}}
    \caption{Photograph of the Ni target mounted in the chamber (A) before deposition and (B) after deposition. A red guide laser is on in both images. Before deposition, the laser is observable from the camera due to diffuse reflections from the surface before melting. After deposition, the guide beam is not visible due to the increased specularity of the melt pool. (C) Deposition rate measured by QCM versus inverse laser power (symbols) and a fit line. Present data (red circles) shown along with data by Ref.~\onlinecite{Smart:2023_2} (green triangles). The present data follow an Arrhenius relationship: Rate [\AA/s] $= 4064 \exp(-1673/\text{P[W]})$. (D) Measured, unscaled deposition rate versus time from QCM for two different depositions. The standard deviation relative to the mean of deposition 1 (deposition 2) is 3.8\% (2.7\%).}
    \label{fig:Target}
\end{figure}

We next present qualitative features of the deposition process. \Cref{fig:ni-before,fig:ni-after} show images of the Ni target before and after deposition, respectively. The remnants of the melt pool formed during deposition can be observed in \Cref{fig:ni-after}. The deposition rate measured by QCM versus inverse laser power is shown in \Cref{fig:Deprate}. To account for the geometric factor from the positioning of the QCM relative to the sample, the measured rates were scaled to match the measured film thickness. The deposition rate exhibits the expected Arrhenius dependence on laser power \cite{Smart:2021,Smart:2023_2}, with a power of 202 W yielding a deposition rate of 0.98 \AA/sec.

During deposition, flux instabilities (on second timescales) and slower variations (minute timescales) were observed, as shown in \cref{fig:Fluctuations}. The slower variations, which manifest here as a downward trend, have been previously reported and are discussed in Refs.~\onlinecite{Smart:2023_2,Smart:2024}. These flux variations might be attributable to a change in surface morphology induced by melting and evaporation and a decrease in the mirror reflectivity as material is deposited on the mirror. Possible mitigation strategies include rastering and defocusing the beam to increase the effective evaporation area at the target surface. \cite{Smart:2023_2,Smart:2024} We did not employ either technique, resulting in a flux variation of up to 10\% over any run. The flux instabilities were on the order of 2-4\% (as measured by QCM), the origin of which is not clear at present.


We note that the observed deposition rate was lower than that previously published in Ref.~\onlinecite{Smart:2023_2}, which concluded that higher energy densities will increase the deposition rate. The source dimensions were also different, with the referenced data set using a 12.7 mm diameter source with a height of 4 mm. However, a smaller source diameter yields lower energy losses through radiation, and thus it should be expected that a higher deposition rate would be achieved, the opposite of what was observed. Our laser was also focused tighter than it was in the reference, which should also increase the flux. Because of these factors and the observed discrepancy in flux being larger than the $\sim25\%$ estimated from the working distance, it is likely that the assumption that conduction does not play a major role is invalid in our setup. In our setup, the target is pressed into the holder, and so conductive heat loss from the target is sufficient to heat the holder to well above 1000~\unit{\celsius}~as indicated by the orange color of blackbody radiation. Because conductive heat loss is linear while radiation and evaporative loss vary more strongly with temperature difference, conductive loss will be most pronounced at lower temperatures, which may be the reason the difference between data sets is most significant at lower powers. This finding highlights the necessity of minimizing thermal contact if the efficiency of evaporation at low powers is a concern.

\begin{figure}
    \centering{
        \phantomsubcaption\label{fig:afm-small}
        \phantomsubcaption\label{fig:afm-large}
        \includegraphics[width=\textwidth]{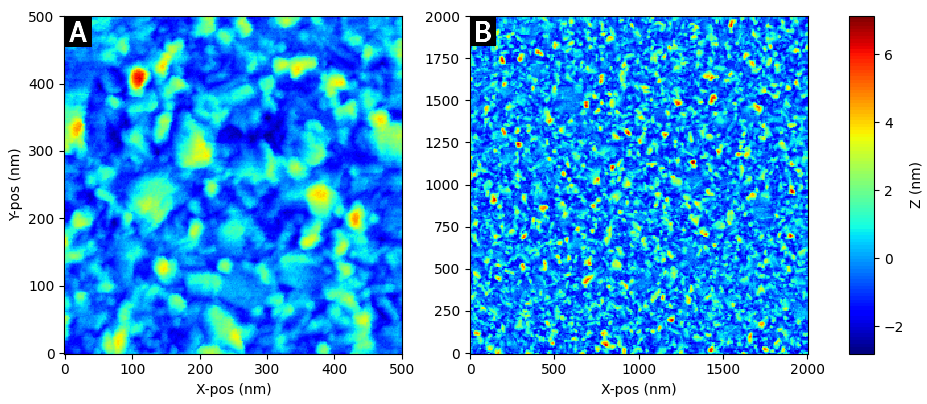}}
    \caption{AFM scan of the film surface. (A) $500 \times 500$ nm scan (256 lines, 0.3 Hz) and (B) $2 \times 2$ \unit{\micro\meter} scan (256 lines, 0.5 Hz). The RMS roughness $R_q = 1.10\pm0.14$ nm.}
    \label{fig:AFM}
\end{figure}

Next, we examine the physical characteristics and surface morphology of the films. The film thickness was measured using ex-situ ellipsometry on a J.A. Woollam M2000 with the built-in profiles for sapphire and Ni. The measurement yielded a value of $14.7\pm0.1$ nm, which was not changed by incorporating roughness or native oxide. This value is quantitatively consistent with step height measurement via atomic-force microscopy (AFM) performed on a step formed by the substrate holder. Although no substrate rotation was used, the thickness variation across the sample should be limited to 0.2\% given the expected cosine flux distribution, which is less than the measurement uncertainty.\cite{Smart:2023_2} The film surface morphology was characterized using a Bruker Dimension Icon AFM in PeakForce Tapping mode with a ScanAsyst-Air probe. The AFM scans of the sample are presented in \Cref{fig:AFM}. We measured the surface morphology at small and large scales in \cref{fig:afm-small} and \cref{fig:afm-large}, respectively, and the same surface morphology is observed at both scales. These scans indicate that the sample exhibits the smoothness expected of a thermally evaporated thin metal film growing by nucleation and coalescence, with RMS roughness $R_q = 1.10\pm0.14$ nm. \cite{Pashley01071965}

\begin{figure}
    \centering{
        \phantomsubcaption\label{fig:XPS_surface}
        \phantomsubcaption\label{fig:XPS_depth}
        \includegraphics[width=\textwidth]{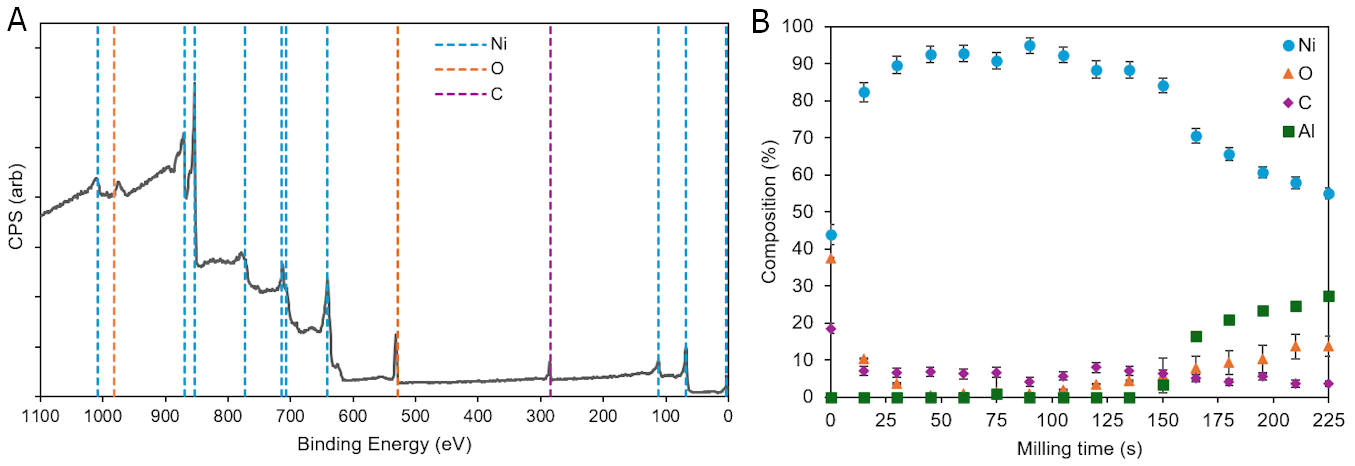}}
    \caption{(A) XPS spectrum of the Ni film surface. O, C, and Ni peak locations are shown according to the Kratos element library. (B) XPS compositional analysis with depth profiling. Detected elements include Ni (blue circles), O (orange triangles), C (magenta diamonds), and Al (green squares). After 150 s of milling the film becomes thin enough that the underlying sapphire substrate begins to appear in the XPS signal. Error bars represent the area uncertainty reported by CASA XPS.}
    \label{fig:XPS}
\end{figure}

The chemical composition of the film was characterized using x-ray photoelectron spectroscopy (XPS). Although not as sensitive as other techniques like secondary ion mass spectrometry, XPS still provides confirmation of the presence of metallic Ni and bounds the O, C, and other impurity content in the film. The measurement was performed using a Kratos Axis Ultra XPS using an Al K $\alpha$ source and used an ion mill for depth profiling. The surface XPS spectrum in shown in \cref{fig:XPS_surface}, and the composition throughout depth profiling in \cref{fig:XPS_depth}. The detected elements were Ni with some amount of O and C. The O content ranges from approximately 0.6\% to 2\% in the bulk, while the C content ranges from about 7\% to 12\%. We attribute the C content to the adventitious C that remained on the sapphire substrate prior to deposition, which diffused into the film as a result of indirect heating of the substrate by the hot target during the deposition process. This adventitious C may also be responsible for a portion of the O content. No other elements were found within the detection limit of $\sim 0.2\%$, indicating that no metallic contamination from the target holder occurred. In a future work, secondary ion mass spectrometry will be employed to better characterize the concentrations of trace impurities, and more aggressive cleaning of the sapphire substrates (c.f. piranha wet clean) will be used to eliminate residual adventitious carbon.

\begin{figure}
    \centering{
        \includegraphics[width=3.47in]{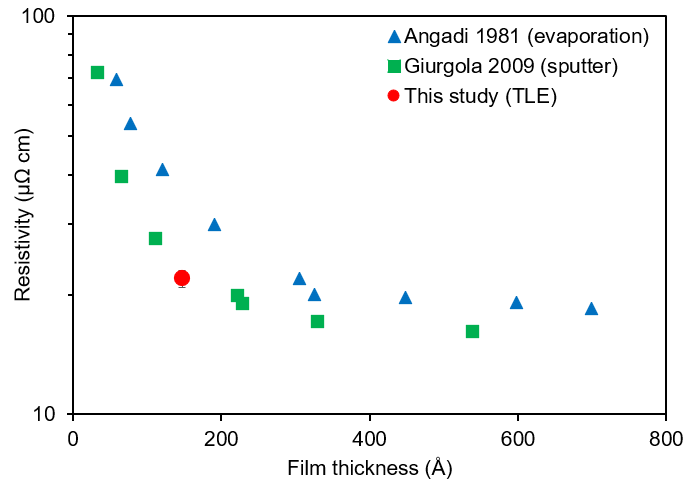}}
    \caption{Electrical resistivity versus film thickness at room temperature. The measurement from this work is indicated as the red point (thickness of 147 \AA), and the uncertainty is smaller than the symbol. Referenced data from Ref.~\onlinecite{Angadi:1981} (blue triangles) and Ref.~\onlinecite{Giurgola:2009} (green squares).}
    \label{fig:resistivity}
\end{figure}

Finally, the electrical resistivity was measured using the van der Pauw method (vdP). The resistivity of thin films is well-known to be a sensitive probe of the structural quality and chemical purity of electrical conductors, and data is readily available in the literature for Ni. Wirebonding was performed using Al wires with a Westbond 7476D Wire Bonder, and the vdP measurement was performed using a Keithley 2400 SourceMeter. At room temperature, the value was measured as $22.0\pm0.2$ $\mu \Omega$cm. \Cref{fig:resistivity} plots this value along with other values for Ni films of varying thickness from prior studies. It is observed that the measured value is compatible with those reported for films of a similar thickness by Ref.~\onlinecite{Giurgola:2009}, which utilized magnetron sputtering. The resistivity is approximately 33\% lower than the best measurement of thermally evaporated ultra-thin Ni by Ref.~\onlinecite{Angadi:1981}. The low resistivity may be due to a lower concentration of electrically-active impurities, grain size, or grain orientation. We conclude that the deposited films are of comparable quality to films deposited using conventional PVD techniques like e-beam evaporation and sputtering. Future work will characterize the temperature-dependence of the resistivity, the residual resistance ratio for films grown via multiple PVD methods, film crystallinity, and the resistivity dependence on growth rate. These measurements will provide further insight into the structural quality and chemical purity and may provide insights into the high electrical conductivity.


In summary, we have reported the deposition and characterization of ultrathin Ni films using a home-built thermal laser evaporation system. Key details of the system design are provided, along with analysis of roughness, composition, and resistivity of non-epitaxial Ni films deposited on sapphire substrates. The films exhibit smoothness consistent with that expected of thin metallic films and electrical resistivity on par with those of films deposited by other PVD techniques. Our work adds to the growing consensus regarding the promising potential of TLE for deposition of high-quality films and compounds, molecular beam epitaxy, and related applications.

\section*{Acknowledgements}
The TLE system was acquired under AFOSR Award FA9550-23-1-0731 and an award from the De Logi Foundation at Caltech. D.S.C. and A.J.M. were supported by AFOSR Award FA9550-22-1-0286. Y.Y. was supported by an Explorer Grant from the Resnick Sustainability Institute at Caltech. F.B.D. acknowledges support from the National Science Foundation Graduate Research Fellowship Program. This material is based upon work supported by the National Science Foundation Graduate Research Fellowship Program under Grant No. 2139433. Any opinions, findings, and conclusions or recommendations expressed in this material are those of the author(s) and do not necessarily reflect the views of the National Science Foundation. We gratefully acknowledge the critical support and infrastructure provided for this work by the Kavli Nanoscience Institute and the Molecular Materials Research Center of the Beckman Institute at the California Institute of Technology for the use of their facilities.

\section*{Data Availability Statement}
The data that support the findings of this study are available from the corresponding author upon reasonable request.

\section*{Conflict of Interest}
The authors have no conflicts to disclose.

\bibliography{bib}

\end{document}